\documentclass[runningheads]{llncs}
\usepackage[utf8]{inputenc}
\usepackage[T1]{fontenc}
\usepackage{graphicx}
\usepackage{subcaption}
\usepackage{amsmath}
\usepackage{amssymb}
\usepackage{xcolor}
\usepackage{cite}

\begin{document}
\title{Resource-aware Computation-Communication Overlap for multi-GPU ML Workloads}
\titlerunning{Resource-aware Computation-Communication Overlap}
%
\author{Minyu Cui\and Miquel Pericàs}
\authorrunning{M. Cui et al.}
%
\institute
{
Chalmers University of Technology and \\ University of Gothenburg, Gothenburg, Sweden \\
%
\email{\{minyu, miquelp\}@chalmers.se}
}
\maketitle              
\begin{abstract}

The rapid growth of large-scale machine learning (ML) has made distributed training across multiple GPUs a fundamental component of modern ML systems. As model sizes and computational throughput continue to increase, communication overhead has become a dominant bottleneck in multi-GPU training, particularly when computation and communication are executed sequentially.
This work explores concurrent execution of computation and collective communication using two portable runtime controls: shared-memory–driven occupancy shaping for computation kernels and elevated scheduling priority for communication kernels. Our approach regulates computation-kernel residency through per-block shared-memory allocation, leaving sufficient on-chip resources for communication kernels to make progress. In addition, assigning higher priority to communication streams ensures steady communication progress once resources become available.
Experiments on NVIDIA A40, A100, H100, and AMD MI250X GPUs demonstrate that the proposed method enables effective computation–communication overlap and reduces total execution time by up to 25.5\%, without modifying vendor libraries or kernel implementations.

\end{abstract}
\section{Introduction}

The computational demands of modern ML models continue to grow rapidly, driven by increasing model sizes and longer sequence lengths during training. 
For example, Meta recently unveiled Llama 4 Behemoth~\cite{hyperparameter_llama4}, with up to 2 trillion parameters. At this scale, distributed execution across multiple GPUs is essential, and collective communications (e.g., allreduce and alltoall) can account for a substantial portion of the total runtime. This bottleneck is especially visible in transformer-based model training, where compute-heavy matrix multiplications and bandwidth-intensive synchronization must both be sustained~\cite{ISCA_Klenk,ISPASS_Diksha,10289240Pati}. 
As computational throughput has outpaced interconnect bandwidth, minimizing communication stalls through computation--communication overlap has become essential for scalable training.
 
Current ML frameworks typically decouple computation and communication by executing them as separate GPU kernels or runtime phases, with limited coordination. While this modular design simplifies development, it often leads to partial serialization and underutilized GPU resources. In modern large language model training, communication can account for up to half of the total execution time~\cite{Comet2025}, making effective overlap critical.
Existing overlap techniques include concurrent execution of computation and communication kernels~\cite{Agrawal2025,ASPLOS24_Chen,Jangda_ASPLOS22,eurosys26,ASPLOS2023_Wang} and combining them into fused kernels~\cite{T3_Pati_ASPLOS,Punniyamurthy_SC2024,Comet2025,FLUX2024}. However, concurrent kernels compete for compute units, cache, and memory bandwidth, and the resulting interference often diminishes the realized overlap~\cite{Agrawal2025}. Decomposition-based strategies mitigate contention by splitting operators into finer-grained pieces, but at the cost of additional kernel launch and synchronization overheads~\cite{ASPLOS24_Chen}. Many methods require intrusive hardware/software changes or specialized runtime features~\cite{Agrawal2025,Jangda_ASPLOS22,T3_Pati_ASPLOS,Punniyamurthy_SC2024,ASPLOS2023_Wang}; others depend on domain-specific languages or custom code generation, which complicates adoption in mainstream frameworks~\cite{Jangda_ASPLOS22}. 
These limitations motivate a portable approach that improves overlap without modifying hardware, vendor libraries, or framework internals.
We target computation paired with collective communication as performance-critical patterns in distributed training. Our contributions are:

\begin{itemize}
\item \textbf{Resource-aware and Priority-aware Overlap.}
We combine occupancy shaping through per-block shared-memory allocation with higher scheduling priority for collective communication kernels. This design preserves correctness, avoids kernel modification, and improves communication progress during overlap.
\item \textbf{Cross-vendor Evaluation and Actionable Insights.}
We evaluate the approach on NVIDIA A40, A100, H100, and AMD MI250X across four representative workloads combining compute-bound and memory-bound GEMM with allreduce and alltoall collectives. Priority-aware overlap consistently improves upon the baseline and reduces execution time by up to 25.5\%. The experimental analysis characterizes the interactions among resource allocation, scheduling priority, and communication progress, providing actionable guidelines for portable overlap in multi-GPU ML workloads.

\end{itemize}

In the remainder of this paper, Section~\ref{background} introduces background, Section~\ref{our_approach_overlap} describes the proposed approach, Section~\ref{ER} presents experimental results, Section~\ref{RW} reviews related work, followed by the conclusion in Section~\ref{Conclusion}.

\section{Background}
\label{background}

The performance of distributed GPU training is largely determined by the interaction between compute-intensive GEMM kernels and bandwidth-intensive collective communication.
\begin{enumerate}
\item[1,] \textbf{GEMM in ML Workloads.}
Linear layers in transformer-based models are commonly implemented as GEMM operations, which account for a large share of training and inference time. A standard GEMM computes $C = \alpha \cdot \text{op}(A) \cdot \text{op}(B) + \beta \cdot C$, where $A \in \mathbb{R}^{M\times K}$, $B \in \mathbb{R}^{K \times N}$, $C \in \mathbb{R}^{M \times N}$. Optimized implementations exploit tiling and shared-memory reuse to maximize arithmetic intensity and GPU occupancy. In distributed settings, GEMM kernels execute repeatedly as part of iterative training, and their effective performance is increasingly shaped by the cost of synchronizing intermediate results across devices.
\item[2,] \textbf{Multi-GPU Collective Communication.}
Multi-GPU systems distribute computation across devices using data or model parallelism. Collective operations such as allreduce, allgather, and alltoall are used to synchronize gradients, exchange activations, and aggregate parameters across GPUs through highly optimized libraries like NCCL and RCCL. For example, in data-parallel training, GEMM kernels compute local gradients, which are then aggregated across all GPUs using allreduce; similarly, expert-parallel workloads rely on alltoall to route tokens between devices. This inter-GPU communication can become a significant bottleneck. Because the GEMM--collective sequences recur at every training iteration, even modest per-iteration communication overhead can accumulate into a substantial end-to-end cost. Consequently, we focus on optimizing GEMM--allreduce and GEMM--alltoall patterns as representative targets in this work.
\item[3,] \textbf{Shared Memory in GPU Architectures.}
Shared memory is a key on-chip resource for GEMM and directly affects occupancy. Increasing shared-memory allocation per thread block reduces the number of blocks that can reside concurrently on a compute unit, while smaller allocations allow higher residency. We exploit this relationship to leave controlled resource slack for communication kernels without modifying kernel logic.
\end{enumerate}

\section{Resource-aware Computation-Communication Overlap}\label{our_approach_overlap}

\begin{figure}[t]
\centering
\begin{subfigure}[b]{0.32\linewidth}
    \includegraphics[width=\textwidth]{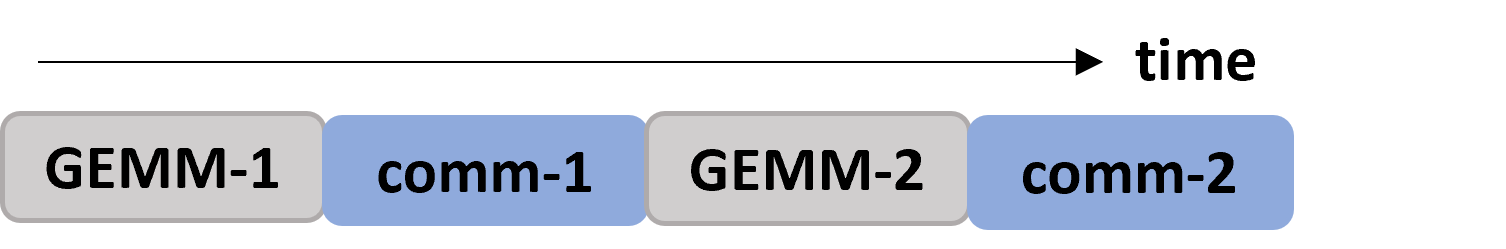}
    \caption{Sequential}
    \label{seq_execution}
\end{subfigure}%
\hspace{2pt}%
\begin{subfigure}[b]{0.32\linewidth}
    \includegraphics[width=\textwidth]{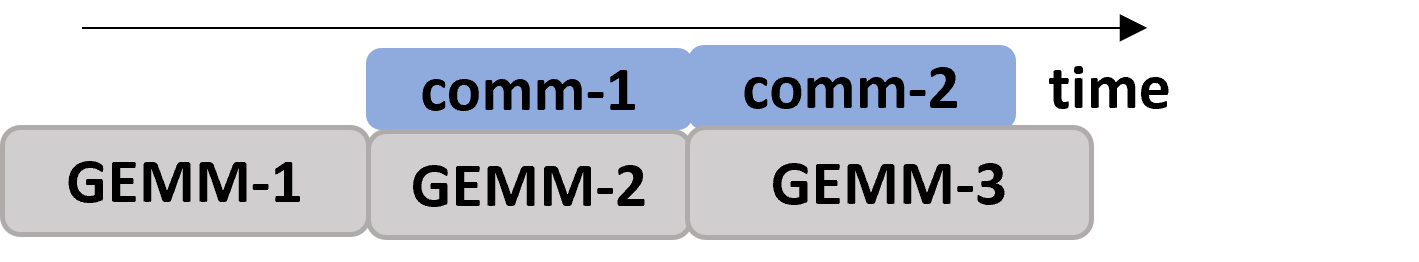}
    \caption{Proposed overlapping}
    \label{overlap_execution}
\end{subfigure}%
\hfill
\begin{subfigure}[b]{0.55\linewidth}
    \includegraphics[width=\textwidth]{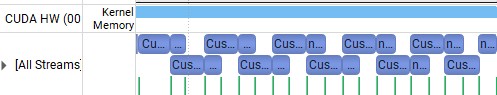}
    \caption{Overlapped \texttt{mb-ar}}
    \label{overlap_illustaretion_SP_mb_A100}
\end{subfigure}
\caption{Overview of proposed overlapped execution}
\label{proposal_overview}
\vspace{-10pt}
\end{figure}

To address the limitations of complex hardware/software modifications and resource contention associated with fine-grained overlap, this section presents a lightweight, portable, resource-aware approach for overlapping computation and communication. 
Our goal is to transform the conventional sequential execution of computation followed by communication into an overlapped execution pattern, as illustrated in Figs.~\ref{seq_execution} and~\ref{overlap_execution}. 
Fig.~\ref{overlap_illustaretion_SP_mb_A100} demonstrates the practical application of the proposed overlap strategy using \texttt{mb-ar} workload described in Section~\ref{two_ML_workloads}. 
In Fig.~\ref{overlap_illustaretion_SP_mb_A100}, longer rectangles represent computation operations, while smaller rectangles denote communication operations, visually highlighting how communication is effectively overlapped with computation.
As NVIDIA and AMD GPUs employ different terminologies for analogous architectural concepts, we adopt CUDA terminology throughout the remainder of this paper for clarity, except where AMD-specific distinctions are required.

\subsection{Resource Residency Control via Shared Memory Allocation}\label{RR_SHMEM}

GPU occupancy is limited by registers, shared memory (or local data share on AMD), and the maximum number of resident thread blocks per compute unit. Communication kernels are intentionally lightweight and low-occupancy~\cite{Nccl_collectiven}, while GEMM kernels often aggressively claim substantial shared memory to maximize data reuse. We use that asymmetry to shape residency: by tuning the GEMM tile sizes, we exploit the amount of shared memory allocated per block and therefore the number of resident GEMM blocks.

For a tiled GEMM kernel, the shared memory allocated per block, $S\_{blk}$, is proportional to $TILE_M \cdot TILE_K + TILE_K \cdot TILE_N$.
By adjusting $TILE_M$, $TILE_N$, and $TILE_K$, we limit occupancy without changing kernel logic, leaving enough SM resources for communication kernels to co-reside with GEMM. This resource control is architecture-agnostic and forms the foundation for the scheduling strategies described in the following sections, enabling effective overlap of computation and communication.

\subsection{Baseline Overlap with Multi-Stream Scheduling}\label{MS-schedule}

To overlap computation with communication, we assign each device two GPU streams~\cite{nvidia_cuda_guide}: a compute stream for GEMM kernels and a communication stream for collectives. Inter-stream dependencies are enforced through lightweight CUDA (or HIP) events: a communication kernel is launched only after its corresponding GEMM kernel signals completion, while subsequent GEMM kernels may execute concurrently with earlier communication. This event-driven, pipelined schedule is portable across vendors, requires no modification to NCCL/RCCL or GEMM kernel code, and exposes concurrency whenever the GPU runtime can co-schedule both kernel types on the available compute units.

\subsection{Optimized Overlap with Higher Scheduling Priority for Collective Communication}\label{optimized_schedule_priority}

GPU architectures rely on internal hardware schedulers to dynamically allocate compute and memory resources among concurrently executing kernels. While this design generally achieves high utilization, it can lead to performance imbalances in heterogeneous workloads where computation-intensive GEMM kernels co-exist with latency-sensitive communication kernels. 
Under the baseline overlap mechanisms of multi-stream scheduling and resource residency control, we observe that GEMM kernels from different iterations tend to be scheduled first. As a result, the GPU scheduler may allocate the majority of resources to these kernels, potentially starving collective communication kernels and increasing overall iteration latency. This observation is consistent with recent work~\cite{Agrawal2025}. 

To address this limitation, we extend our baseline overlap mechanisms with an optimized strategy that leverages GPU stream scheduling priorities. 
Specifically, we assign communication kernels a higher scheduling priority than GEMM kernels from overlapping iterations. 
This prioritization ensures that communication operations can make forward progress whenever resources become available, while still respecting all intra-iteration dependencies. This effectively overlaps data transfers with matrix multiplication, which is critical for sustaining high throughput in pipelined iterative workloads.
Formally, let $K_g^i$ and $K_c^i$ denote the GEMM and the collective communication kernel of iteration $i$. Correctness requires the dependency $K_g^i \to K_c^i$. Across iterations, we explicitly allow the scheduler to prioritize $K_c^{i} \succ K_g^{i+1}$ where $\succ$ denotes higher scheduling priority specified by \textit{cudaStreamCreateWithPriority} or \textit{hipStreamCreateWithPriority}.
The prioritization affects only inter-iteration concurrency, allowing communication kernels from earlier iteration to run concurrently with computation kernels from later iterations when possible, without violating correctness or data consistency.
Our evaluation demonstrates that this priority-aware scheduling approach improves overall execution efficiency. We present experimental results illustrating these benefits in the following section.

\section{Experimental Results}\label{ER}

This section evaluates our approach on four modern multi-GPU platforms using representative ML workloads. Due to page limitations, we present a subset of results that illustrate key trends. The complete set of results across all configurations is consistent with the observations reported here, and any notable deviations are explicitly discussed. After confirming that the baseline overlap achieves effective concurrency, we examine the optimized, priority-aware overlap and analyze its impact on execution time, kernel concurrency, and the degree of communication–computation overlap across architectures and workloads.

\subsection{Experimental Setup}

Table~\ref{tbl:setup} provides an overview of the hardware platforms, workload configurations, and tile size settings employed in our evaluation.

\subsubsection{Multi-GPU Platforms.}\label{GPU_platforms}

The NVIDIA-based platforms (A40, A100, and H100) each contain four GPUs, while the AMD-based platform consists of eight MI250X GPUs. These counts reflect the standard single-node configuration of each testbed. Since our overlap mechanism operates at the per-device level (via occupancy shaping and stream priorities), GPU count affects only the collective scale, which we fix at 896 MB across platforms to keep the comparison controlled.
Together, these systems represent a diverse set of architectural designs, including NVIDIA’s Ampere and Hopper families and AMD’s CDNA2 architecture, providing a representative testbed for evaluating our approach.

\begin{table}[t]
\scriptsize
\centering
\caption{Summary of platforms, workloads, and tile configurations.}
\label{tbl:setup}
\vspace{2pt}
\setlength{\tabcolsep}{3pt}
\renewcommand{\arraystretch}{0.95}

{\bfseries Platforms}\\[1pt]
\begin{tabular}{lcccc}
\hline
 & A40 & A100 & H100 & MI250X \\ \hline
\#GPUs    & 4      & 4      & 4      & 8      \\
\#SMs/CUs & 84     & 108    & 132    & 110    \\
L2        & 6 MB   & 40 MB  & 50 MB  & 8 MB   \\
HBM       & 48 GB  & 40 GB  & 80 GB  & 64 GB  \\
L1+SMEM   & 128 KB & 192 KB & 256 KB & 80 KB  \\ \hline
\end{tabular}
\\[4pt]

{\bfseries Workloads and tile configurations}\\[1pt]
\begin{tabular}{lcc|lccc}
\hline
\multicolumn{3}{c|}{Workloads} & \multicolumn{4}{c}{Tile configurations} \\
 & GEMM ($M{\times}N{\times}K$) & Coll.\ data & & $TILE_m$ & $TILE_n$ & $TILE_k$ \\ \hline
\texttt{cb-ar}  & $8192{\times}8192{\times}8192$  & 896 MB & \texttt{opt1} & 64 & 64 & 32 \\
\texttt{mb-ar}  & $8192{\times}57344{\times}8192$ & 896 MB & \texttt{opt2} & 64 & 64 & 64 \\
\texttt{cb-a2a} & $8192{\times}8192{\times}8192$  & 896 MB &               &    &    &    \\
\texttt{mb-a2a} & $8192{\times}57344{\times}8192$ & 896 MB &               &    &    &    \\ \hline
\end{tabular}

\renewcommand{\arraystretch}{1}
\vspace{-8pt}
\end{table}

\subsubsection{Representative ML Workloads.}\label{two_ML_workloads}

The evaluated GEMM dimensions are derived from LLaMA model training~\cite{hyperparameter_llama3} with a sequence length of 8192 tokens, capturing realistic compute patterns of large-scale transformer-based models. Following prior work~\cite{Agrawal2025}, we pair two GEMM categories, compute-bound (\texttt{cb}) and memory-bound (\texttt{mb}), with two collective communication operations, each transferring 896~MB of data. 
For allreduce (\texttt{ar}), we construct \texttt{cb-ar} and \texttt{mb-ar} configurations to represent computation-dominant and communication-dominant training phases, respectively. 
For alltoall (\texttt{a2a}), we model MoE-style expert-parallel communication patterns using the identical GEMM dimensions, resulting in \texttt{cb-a2a} and \texttt{mb-a2a} configurations.  

\subsubsection{GEMM Tile-size Configurations.}\label{two_tile_configuration}

We test two tile-size configurations: \texttt{opt1} ($TILE_k{=}32$) and \texttt{opt2} ($TILE_k{=}64$), both with $TILE_m{=}TILE_n{=}64$. The larger tile size increases arithmetic intensity of GEMM but also increases shared-memory usage. This setup allows us to study the trade-off between GEMM performance potential and communication–computation overlap opportunity. 

For each experiment, the first one-third of runs are treated as warm-up and excluded from analysis; the remaining runs are used for measurement. Each reported value is the mean across the runs. The observed run-to-run variance is small and does not affect the qualitative trends we discuss.
We first evaluate the effectiveness of the proposed resource-aware overlap strategy and then analyze how GEMM tile-size configurations interact with overlap behavior.

\subsection{Resource-aware Computation-Communication Overlap}\label{optimized_overlap}

\begin{figure*}[t]
\centering
{\includegraphics[width=0.8\textwidth]{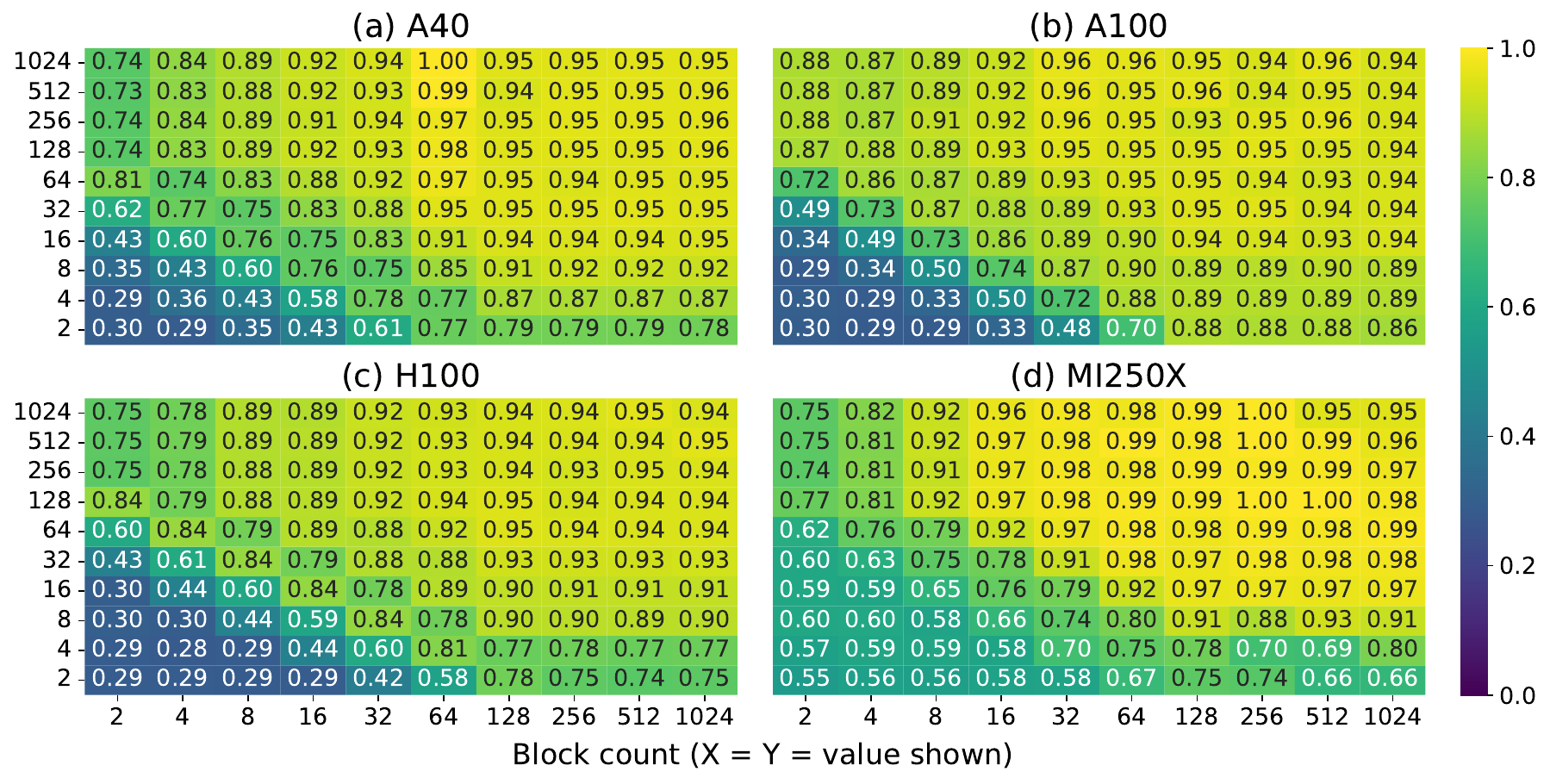}
\label{timeratio_mb1_896MB_A40}}
\vspace{-4mm}   
\caption{TimeRatio of baseline overlap for \texttt{cb-ar}.}
\label{timeration_baseline_opt1_cmb1_896MB}
\vspace{-10pt}
\end{figure*}

\subsubsection{Baseline Computation-Communication Overlap}

We begin by confirming that the baseline implementation enables
effective computation–communication overlap across the four evaluated platforms. To quantify overlap effectiveness, we define:
 $Time Ratio = \frac{t_{overlap} }{t_{sequential}}$,
where $t_{sequential}$ is the execution time when computation and communication are executed sequentially, and $t_{overlap}$ is the execution time when overlap is enabled.
A lower ratio indicates more effective overlap. Across all platforms, the baseline mechanism is most effective when thread-block counts are highly constrained to leave sufficient resource slack for kernel co-residency, as shown in Fig.~\ref{timeration_baseline_opt1_cmb1_896MB}. In this regime, $TimeRatio$ drops to approximately 0.3. However, such configurations generally do not yield optimal GEMM performance, as the reduced number of resident thread-blocks limits parallelism and lowers compute unit utilization. 
As the thread-block count increases, pressure on shared memory, registers, and warp slots reduces the scheduler’s ability to co-locate GEMM and communication kernels. Consequently, the overlapped execution time converges toward sequential execution ($TimeRatio \approx 1$). 
These results highlight that effective overlap requires sufficient resource slack to enable kernel co-residency, which depends on the interplay between thread-block configuration, on-chip resource contention, and hardware scheduling decisions. Among the four platforms, MI250X shows the weakest benefit. In some cases of \texttt{mb-ar} (not shown here), the overlapped execution even becomes slightly slower than sequential execution.
This behavior likely stems from the smaller per-CU shared memory and L2 cache capacity on MI250X, which leave limited resources for kernel co-residency.

Overall, these results confirm that the baseline mechanism can enable overlap but also expose its key limitation: communication progress becomes fragile under resource contention. To address this limitation, we next evaluate the optimized, priority-aware strategy.
We report two metrics: normalized execution time (referred as Norm.\ time), and overlap rate, defined as the fraction of communication that executes concurrently with computation.

\subsubsection{Priority-aware Scheduling across Platforms and Workloads.}\label{effect_optimized_overlap}

\begin{figure*}[t]
\centering
{\includegraphics[width=\textwidth]{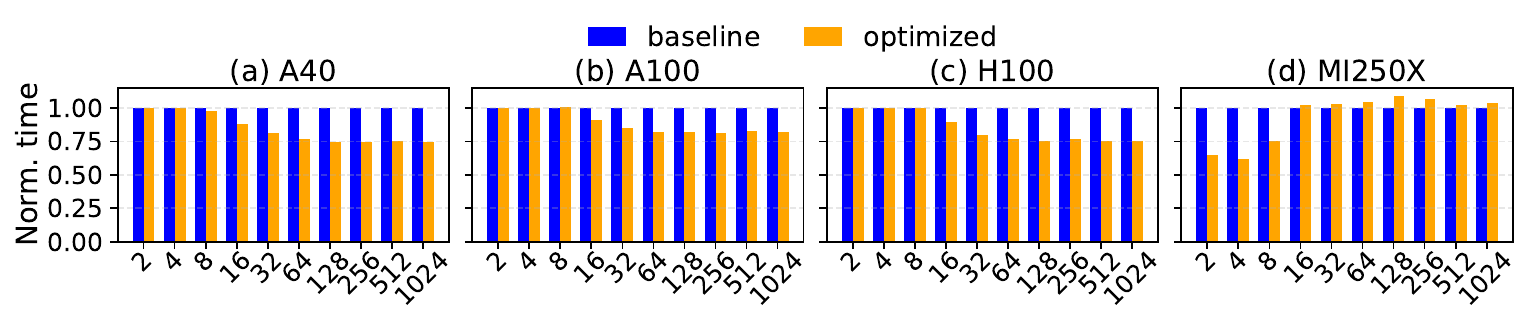}
\label{timeratio_mb1_896MB_A40}}
\vspace{-8mm}   
\caption{Norm. time (optimized overlap normalized
to the baseline) for \texttt{cb-ar}. Block count (X = Y = value shown). This convention applies to subsequent bar plots)}
\label{speedup_optimized_baseline_cb_896MB}
\vspace{-10pt}
\end{figure*}

\begin{figure*}[t]
\centering
\begin{subfigure}[b]{0.24\linewidth}
{\includegraphics[width=1\textwidth]{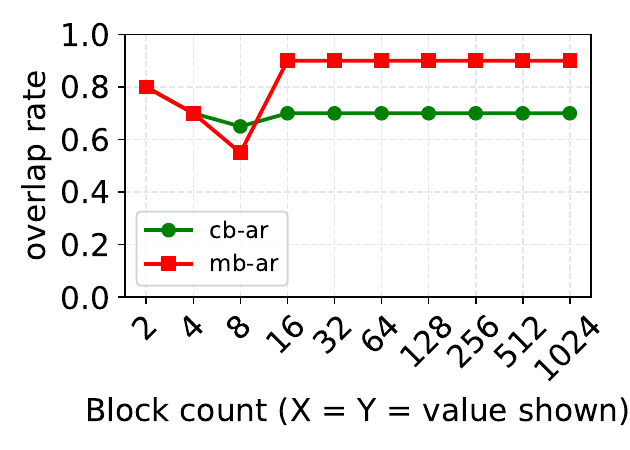}
\caption{A40}
\label{overlap_rate_cb_mb_A40}}
\end{subfigure}
\begin{subfigure}[b]{0.24\linewidth}
{\includegraphics[width=1\textwidth]{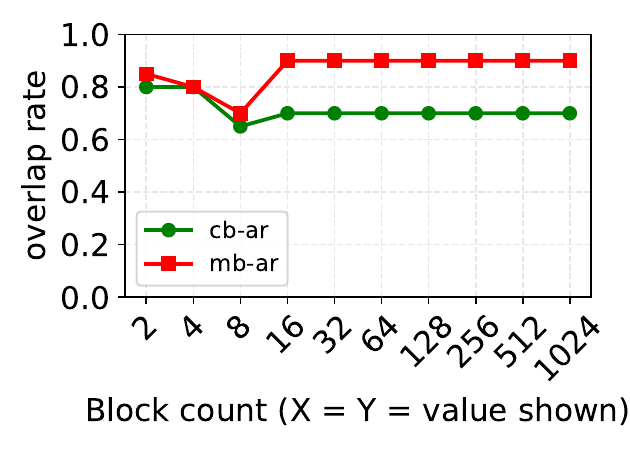}
\caption{A100}
\label{overlap_rate_cb_mb_A100}}
\end{subfigure}
\begin{subfigure}[b]{0.24\linewidth}
{\includegraphics[width=1\textwidth]{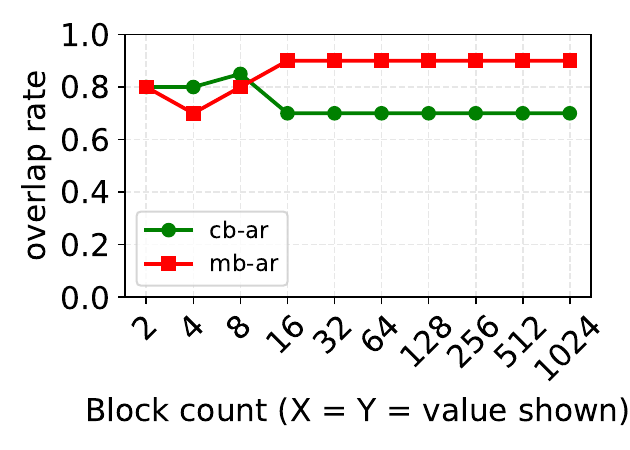}
\caption{H100}
\label{overlap_rate_cb_mb_H100}}
\end{subfigure}
\begin{subfigure}[b]{0.24\linewidth}
{\includegraphics[width=1\textwidth]{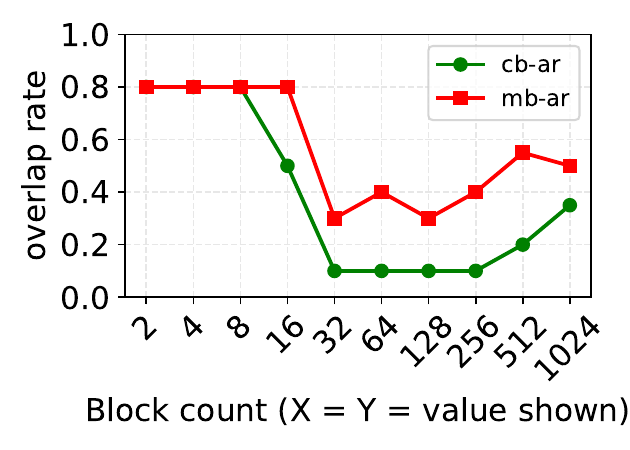}
\caption{MI250X}
\label{overlap_rate_cb_mb_MI250X_SP}}
\end{subfigure}
\vspace{-6pt}
\caption{The overlap rate.}
\label{overlap_rate_cb_mb_896MB}
\end{figure*}

We compare the optimized priority-aware strategy against the baseline across evaluated platforms and workloads in Fig.~\ref{speedup_optimized_baseline_cb_896MB}. Overall, prioritizing communication kernels improves overlap efficiency, although the degree of improvement depends on both GPU architecture and workload characteristics.
On NVIDIA GPUs, the optimized strategy produces notable reductions in total execution time, particularly when the baseline already enables partial concurrency. Taking \texttt{cb-ar} for example, the optimized overlap allows communication kernels to make progress earlier, mitigating the tail effect observed in the baseline. The improvement is especially pronounced at medium to high thread-block counts. When thread-block counts are small, contention is low and the baseline already captures most of the achievable benefit; as the block count increases, however, GEMM kernels compete more aggressively for shared resources, and the baseline may allow them to delay or starve communication progress. 
Prioritizing communication mitigates this interference, enabling higher-priority communication operations to advance steadily even under substantial GEMM activity. As a result, execution time is reduced by up to 25.5\% for \texttt{cb-ar} on A40. Similar trends are observed for the remaining workloads, where the optimized strategy reliably prevents communication from being stalled behind long-running GEMM operations.

On MI250X, the optimized overlap can occasionally underperform  baseline at higher thread-block counts (Fig.3(d)). This is primarily attributed to its multi-chip architecture: each Graphics Compute Die (GCD) offers less per-CU shared memory than its NVIDIA counterparts. As the block count grows, the per-block resource allocation enforced by optimized scheduling saturates shared resources more quickly, reducing block co-residency on a GCD. Although communication kernels receive higher priority, the resulting reduction in GEMM concurrency limits effective overlap, and the loss in computational throughput outweighs the benefit gained from prioritized communication.

Fig.~\ref{overlap_rate_cb_mb_896MB} further quantifies the achieved overlap rate. On NVIDIA platforms, the optimized strategy reaches up to 90\% overlap rate, which represents the practical upper bound, since the intra-iteration dependency $K_g^i \to K_c^i$ forces the last communication kernel to complete only after its corresponding GEMM kernel finishes (seen in Fig.~\ref{overlap_illustaretion_SP_mb_A100}). On MI250X, the overlap rate remains notably lower, consistent with the architectural constraints discussed above.
In summary, prioritizing communication kernels significantly enhances overlap efficiency across diverse platforms and workloads, yielding substantial execution-time reductions. However, the benefits are sensitive to architecture and workload characteristics, with MI250X's design limiting the effectiveness of this strategy at higher block counts.
Next, we examine how tile-size tuning interacts with the effectiveness of the optimized overlap strategy.

\subsection{Impact of GEMM Tile Sizes on Overlap Efficiency}\label{tile_setup_overlap}

\begin{figure*}[t]
\centering
{\includegraphics[width=\textwidth]{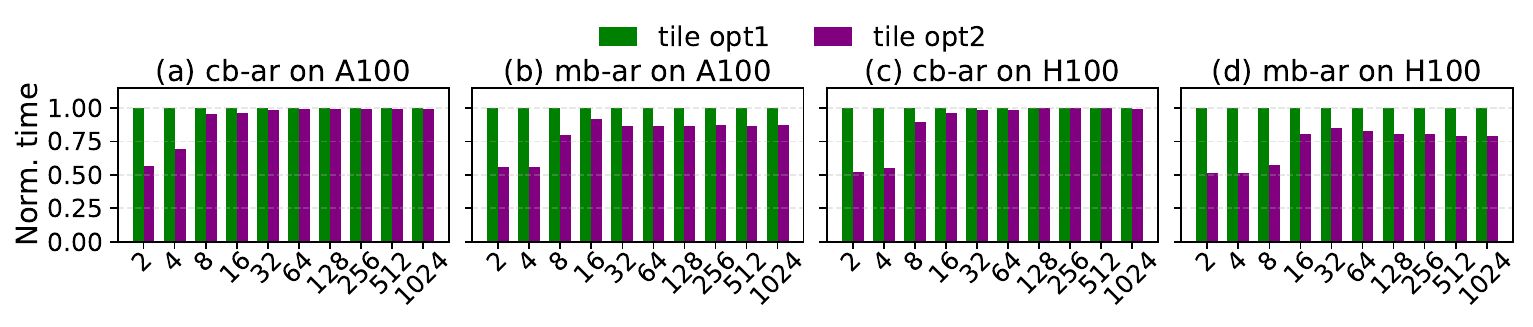}
\label{timeratio_mb1_896MB_A40}}
\vspace{-8mm}   
\caption{Norm. time (tile configuration \texttt{opt2} normalized to \texttt{opt1}).}
\label{optimized_overlap_opt1_opt2_ar}
\vspace{-10pt}
\end{figure*}

\begin{figure*}[t]
\centering
{\includegraphics[width=\textwidth]{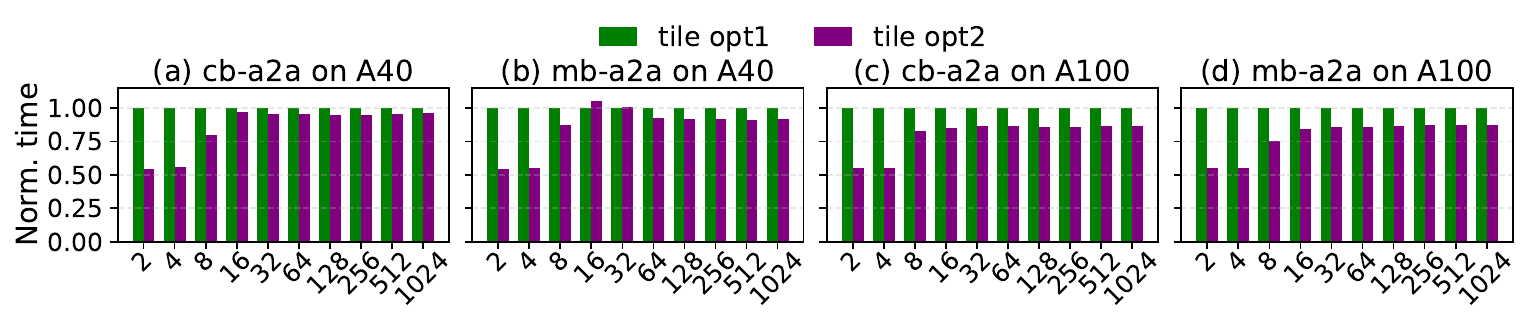}
\label{timeratio_mb1_896MB_A40}}
\vspace{-8mm}   
\caption{Norm. time (tile configuration \texttt{opt2} normalized to \texttt{opt1}).}
\label{optimized_overlap_opt1_opt2_a2a}
\vspace{-10pt}
\end{figure*}
 
In this section, we examine how tile size interacts with the priority-aware overlap strategy. Fig.~\ref{optimized_overlap_opt1_opt2_ar} and~\ref{optimized_overlap_opt1_opt2_a2a} report the execution time of the larger tile configuration (\texttt{opt2}) normalized to \texttt{opt1}. Due to limited platform access, allreduce (\texttt{ar}) comparison is conducted on A100 and H100, while alltoall (\texttt{a2a}) comparison is conducted on A40 and A100.
Overall, \texttt{opt2} achieves lower execution times across a wide range of thread-block counts, indicating that the larger tiles benefit more from communication prioritization when overlap is present. This effect is particularly pronounced for memory-bound workload \texttt{mb-ar}, where the GEMM kernel spends a substantial fraction of time waiting on data. In this case, increasing $TILE_k$ enlarges the amount of computation performed per tile, extending the compute phase. The larger computation window provides additional opportunity for the communication stream, boosted by higher scheduling priority, to make forward progress while GEMM continues in parallel. 
Similar trends hold for \texttt{a2a}: \texttt{opt2} generally matches or outperforms \texttt{opt1}, with the most noticeable gains appearing for \texttt{mb-a2a}. A minor exception occurs for \texttt{mb-a2a} on A40 at thread-block counts of 16 and 32 (Fig.6(b)), where \texttt{opt2} is marginally slower.

The relative advantage of \texttt{opt2} also depends on the thread-block count. At low to moderate counts, the GPU retains sufficient headroom to schedule both GEMM and collective kernels concurrently; the larger $TILE_k$ produces longer computation phases per tile, increasing the compute-to-communicate ratio and enabling the prioritized scheduler to hide communication latency more effectively. At very high block counts, the two tile configurations converge in performance, as occupancy saturation limits the scheduler's ability to exploit prioritization. These results suggest that larger GEMM tiles can moderately enhance the effectiveness of priority-based overlap, particularly for memory-bound workloads. 
However, fully characterizing the interaction between tile size, resource utilization, and priority-aware scheduling requires a broader exploration of the tile-size design space, which we leave for future work.

\section{Related work}\label{RW}

Prior work on computation--communication overlap falls into two broad categories: concurrent kernel scheduling and kernel fusion.

\textbf{Concurrent Kernel Scheduling.}
Several approaches overlap separate computation and communication kernels through scheduling policies or operator decomposition. ConCCL~\cite{Agrawal2025} studies interference between concurrent kernels and offloads communication to DMA engines to reduce contention. Centauri~\cite{ASPLOS24_Chen} decomposes operators at configurable granularity and searches for an efficient communication schedule. NanoFlow~\cite{NanoFlow2024} combines resource partitioning with multi-dataflow scheduling to sustain overlap across pipeline stages. FlashOverlap~\cite{eurosys26} introduces a signaling mechanism that enables interference-free, tile-wise overlap without being tied to specific communication primitives. Compiler-driven methods~\cite{Jangda_ASPLOS22,ASPLOS2023_Wang} transform execution order and generate specialized kernels to coordinate overlap at compile time. Other work~\cite{8821001Runtime} applies runtime concurrency control and operation reordering to reduce inter-kernel interference during training. While effective, these approaches often require custom runtimes, domain-specific compilers, or hardware extensions that limit portability.

\textbf{Kernel Fusion.}
Another line of work fuses communication and computation into a single kernel~\cite{T3_Pati_ASPLOS,Punniyamurthy_SC2024,FLUX2024,Comet2025,TileLink2025}. Fused kernels can eliminate launch overhead and improve data locality, but typically depend on invasive kernel modifications, specialized compilers, or hardware-specific intrinsics. Works targeting MoE models~\cite{Comet2025,FasterMoE2022,Lancet2024} further exploit model-specific dataflow to overlap expert dispatch with computation. These methods achieve strong performance within their target scenarios but are difficult to generalize across vendors and workload types.

In contrast, our approach requires no modifications to vendor libraries, kernel source code, or compiler toolchains. It relies solely on two portable runtime controls, i.e., shared-memory-driven occupancy shaping and stream-priority scheduling, and has been validated on four GPU architectures spanning two vendors.

\section{Conclusion}\label{Conclusion}
We present a lightweight approach for overlapping computation and communication for multi-GPU ML workloads without modifying kernel logic or vendor libraries. The approach uses two portable runtime controls: shared-memory-driven occupancy shaping for GEMM operations and higher scheduling priority for communication kernels. Evaluated on NVIDIA A40, A100, H100, and AMD MI250X across both allreduce and alltoall collective patterns, 
the results show that the key challenge is not merely enabling concurrency, but sustaining communication progress under resource contention. 
The resource-aware baseline provides the necessary kernel co-residency, and priority-aware scheduling design further reduces total execution time, up to 25.5\%, by ensuring that collective operations advance promptly. Future work includes exploring adaptive runtime policies that automatically tune occupancy and priority settings across diverse workloads, and conducting end-to-end distributed training performance evaluations.

%
%
%
%

\end{document}